\documentclass[twocolumn,letter]{jpsj3}
\usepackage{txfonts}
\usepackage{bm}
\usepackage{color}

\title{Switching of Conducting Planes by Partial Dimer Formation in IrTe$_2$}

\author{
\name{Tatsuya \surname{Toriyama}}$^1$, 
\name{Masao \surname{Kobori}}$^1$, 
\name{Takehisa \surname{Konishi}}$^2$, 
\name{Yukinori \surname{Ohta}}$^1$\thanks{E-mail: ohta@faculty.chiba-u.jp}, 
\name{Kunihisa \surname{Sugimoto}}$^3$, 
\name{Jungeun \surname{Kim}}$^3$, 
\name{Akihiko \surname{Fujiwara}}$^3$, 
\name{Sunseng \surname{Pyon}}$^{4,5}$, 
\name{Kazutaka \surname{Kudo}}$^5$, and
\name{Minoru \surname{Nohara}}$^5$\thanks{E-mail: nohara@science.okayama-u.ac.jp}
}

\inst{
$^1$Department of Physics, Chiba University, Chiba 263-8522, Japan \\
$^2$Graduate School of Advanced Integration Science, Chiba University, Chiba 263-8522, Japan \\
$^3$Japan Synchrotron Radiation Research Institute (JASRI), SPring-8, Hyogo 679-5198, Japan \\
$^4$Department of Applied Physics, The University of Tokyo, Tokyo 113-8656, Japan \\
$^5$Department of Physics, Okayama University, Okayama 700-8530, Japan
}

\recdate{September 15, 2013}

\abst{Single-crystal X-ray diffraction was employed to study the structural-electronic phase transition of IrTe$_2$ at approximately 270 K. 
The low-temperature structure was found to be a triclinic (space group $P\bar{1}$) 
characterized by the partial formation of Ir$_2$ dimers in the triangular lattice of IrTe$_2$, 
resulting in a structural modulation with a wave vector of ${\bm q} = (1/5, 0, -1/5)$. 
First-principles band calculations demonstrate that tilted two-dimensional Fermi surfaces emerge in the triclinic phase,
suggesting that switching of the conducting planes occurs from the basal plane of the trigonal IrTe$_2$ to the tilted plane normal to ${\bm q}$ of the triclinic IrTe$_2$. }

\begin{document}
\maketitle

Making and breaking chemical bonds in a solid often results in a drastic change in the physical properties and thus offers useful ways of controlling the electronic states; 
examples includes 
the metal-insulator transition \cite{PhysRevB.70.161102, PhysRevLett.95.196404,PhysRevLett.103.146405, Nature416_155, PhysRevLett.110.166402}, 
superconductivity \cite{PhysRevB.83.060505,JPSJ.80.103701,PhysRevB.86.100505,PhysRevB.85.024525,JPSJ.82.063704,PhysRevB.64.214514}, 
and the magnetic quantum critical point \cite{NaturePhys7_207}. 
For instance, the formation of arsenic dimers that occurs in the adjacent FeAs layers of iron-based superconductors results in a two-dimensional to three-dimensional change in the topology of the Fermi surface, which in turn leads to the loss of the iron magnetic moment and thus the high-transition-temperature superconductivity \cite{PhysRevB.83.060505,JPSJ.80.103701}. 
In this Letter, we report that the partial formation of iridium dimers in the triangular lattice of IrTe$_2$ results in the emergence of dimerized Ir$_2$ and non-dimerized Ir layers that are stacked alternately along a structural modulation vector of ${\bm q} = (1/5, 0, -1/5)$. 
First-principles band calculations demonstrate that the partial formation of iridium dimers results in tilted two-dimensional Fermi surfaces, thus switching the conducting planes from the basal plane of the trigonal IrTe$_2$ at high temperatures to the plane normal to ${\bm q}$ of the triclinic IrTe$_2$ at low temperatures.  

IrTe$_2$ crystallizes in a trigonal CdI$_2$-type structure with the space group $P{\bar 3}m1$ ($\sharp$~164). 
Edge-sharing IrTe$_6$ octahedra form two-dimensional IrTe$_2$ layers that are stacked along the $c$-axis, 
as shown schematically in Figs.~1(a) and 1(b).
In each layer, the Ir atoms form a regular triangular lattice with a uniform Ir-Ir bond length of 3.925 {\AA}.
This compound undergoes a first-order structural phase transition at approximately 270 K \cite{JLowTempPhys117_1129}. 
Matsumoto et al.~\cite{JLowTempPhys117_1129} concluded that the average structure below 270 K is monoclinic with the space group $C2/m$ ($\sharp$~12) in which the Ir-Ir bond length along the monoclinic $b$-axis is uniformly reduced so that the regular triangular lattice is deformed into an isosceles triangular lattice. 
Recently, transmission electron microscopy experiments have revealed that the structural phase transition is accompanied by the evolution of non-sinusoidal structural modulation of ${\bm q} = (1/5, 0, -1/5)$ \cite{PhysRevLett.110.127209}, 
indicating that the structural phase transition is not of a simple charge-density-wave (CDW) type. 
An optical spectroscopy study suggested that the structural phase transition is driven by a reduction in the kinetic energy of the electrons due to Te $5p$ band splitting below the transition temperature \cite{ScientificReports3.1153}. 
X-ray photoemission spectroscopy \cite{PhysRevB.86.014519} and angle-resolved photoemission spectroscopy (ARPES) \cite{JPSJ.82.093704} have 
indicated the importance of the orbital degeneracy of Ir $5d$ and/or Te $5p$ for the transition. 
Very recently, Oh et al. \cite{PhysRevLett.110.127209} suggested that the structural phase transition involves the depolymerization-polymerization of anionic Te bonds. 
Interestingly, suppression of the structural phase transition by chemical doping or intercalation results in the emergence of superconductivity at up to 3.1 K \cite{JPSJ.81.053701,PhysRevLett.108.116402,PhysRevB.87.180501,JPSJ.82.085001,PhysRevB.87.121107}. 
Thus, it is important to determine the low-temperature crystal structure to elucidate the structural-electronic phase transition mechanism of IrTe$_2$. 

Single crystals of IrTe$_2$ were grown using the self-flux technique \cite{ScientificReports3.1153,Pyon2013}. 
A mixture of Ir and Te powder at an atomic ratio of 18:82 was placed in an alumina crucible that was sealed in an evacuated small quartz tube and then placed into a larger quartz tube. The larger quartz tube was evacuated and sealed, and the mixture was initially heated to 950$^\circ$C; this temperature was maintained for 10h. The temperature was then increased slowly to 1160$^\circ$C, before the mixture was cooled to 900$^\circ$C at a rate of 1--2$^\circ$C/h. The quartz tube was then quenched in ice water after decantation of the Te flux. The obtained crystals exhibited a phase transition at approximately 270 K as determined by resistivity and magnetization measurements \cite{Pyon2013}.

Synchrotron radiation X-ray diffraction measurements were performed on beamline BL02B1 at SPring-8, Japan, using 
monochromatized X-ray with an energy of 35.05 keV ($\lambda$ = 0.354 {\AA}). 
Measurements were performed at 300 and 20 K using a cryogenic He-flowing system (XR-HR10KS, Japan Thermal Engineering Co.~Ltd.). 
A cylindrical imaging plate with a camera length of 191.3 mm was adopted \cite{sugimoto}. 
At $T$ = 300 and 20 K, diffraction images in a scattering vector range of up to approximately 34 {\AA}$^{-1}$ were taken 
every 13 and 24 frames, respectively, at an exposure time of 1 min per image. 
The oscillation angle of the crystal ($\omega$) in each frame was 15$^\circ$. 
Typical diffraction images are shown in Fig.~2. 
Data were collected and processed using the RAPID-AUTO program (Rigaku) and were corrected for Lorentz and polarization effects. 
The structures were solved by direct methods (SHELXS-97) and expanded using Fourier techniques. 
All the atoms were refined anisotropically (SHELXL-97). 
The final cycle of the full-matrix least-squares refinement was based on \{all data, $I > 2\sigma(I)$\} = \{261, 261\} and \{11716, 10472\} observed reflections 
for $T$ = 300 and 20 K, respectively. 
Unweight and weighted agreement factors of 
$R = \sum \left| \left| F_0 \right| - \left| F_c \right| \right| / \sum \left| F_0 \right|$, 
$R_1 = \sum \left| \left| F_0 \right| - \left| F_c \right| \right| /  \sum \left| F_0 \right|$  ($F_0 > 4\sigma(F_0)$), 
and 
$wR_2 = [\sum (w(F_0^2 - F_c^2)^2 / \sum w(F_0^2)^2]^{1/2}$ 
were used. 
The $R$, $R_1$, and $wR_2$ values were \{0.0179, 0.0179, 0.0478\} and \{0.0754, 0.0694, 0.242\} for $T$ = 300 and 20 K, respectively. 
The crystal structure of the triclinic phase ($T$ = 20 K) was refined as a twin-domain crystal using the TWIN and BASF commands of SHELXL-97. 
The final composition ratio of the twin-domain crystal was 0.712(1):0.288(1). 
Crystallographic data and fractional coordinates obtained at both temperatures are summarized in Table 1 and 2, respectively.

\begin{figure}
\begin{center}
\includegraphics[width=8cm]{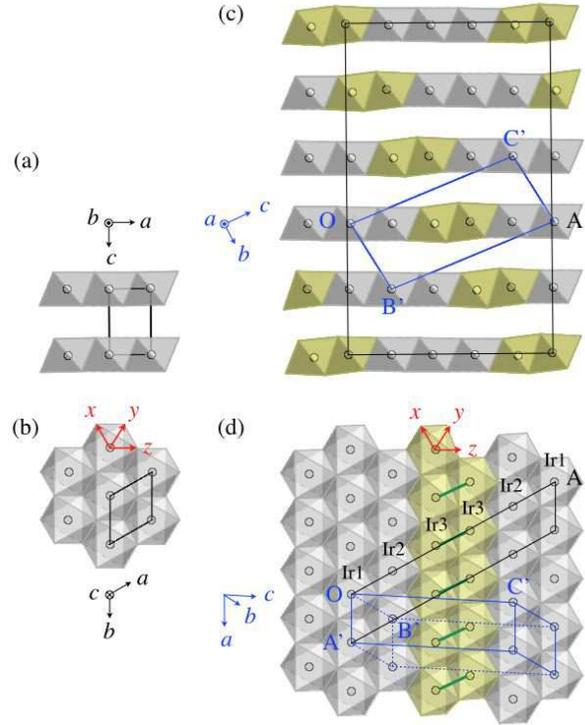}
\caption{
(Color online) (a) and (b) The crystal structure of IrTe$_2$ at 300 K. Black lines represent the unit cell of the trigonal lattice.
(c) and (d) The crystal structure of IrTe$_2$ at 20 K. Blue lines represent the unit cell of the triclinic lattice.  
Black lines represent the $5a \times b \times 5c$ supercell ($a$, $b$, and $c$ are the high-temperature trigonal-cell parameters).
}
\end{center}
\end{figure}

\begin{figure}
\begin{center}
\includegraphics[width=8cm]{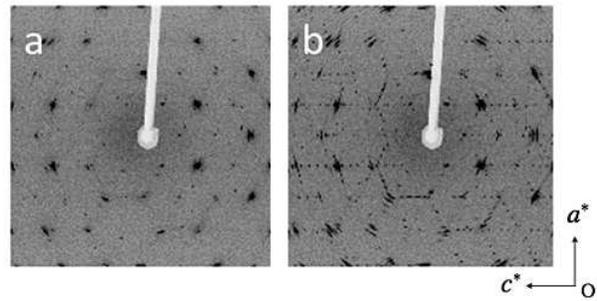}
\caption{
Typical diffraction images of IrTe$_2$ taken at (a) 300 K and (b) 20 K. 
}
\end{center}
\end{figure}

\begin{table}
\caption{
Crystallographic data of IrTe$_2$ at 300 and 20 K. 
}
\begin{tabular}{llll}
\hline
\hline
\multicolumn{1}{l}{Formula} & & \multicolumn{2}{c}{IrTe$_2$} \\
\multicolumn{1}{l}{Formula weight} & & \multicolumn{2}{c}{447.42} \\
\hline
Temperature (K)  & & 300 & 20 \\
Space group & & $P{\bar 3}m1$ ($\sharp$ 164) & $P{\bar 1}$ ($\sharp$ 2) \\
$a$ (\AA) & & 3.92530(10) & 3.94860(10) \\
$b$ (\AA) & & 3.92530(10)  & 6.64480(10) \\
$c$ (\AA) & &  5.39520(10) &  14.4045(3) \\
$\alpha$ ($^\circ$) & &  90 & 80.245(6) \\
$\beta$ ($^\circ$) & & 90  & 87.404(6)  \\
$\gamma$ ($^\circ$) & & 120  & 72.511(5) \\
Volume (\AA$^3$) & &  71.992(3) & 355.252(13) \\
$Z$ & & 1  & 5 \\
$R_{\rm int}$ & &  5.79 & 6.57 \\
\hline
\hline
\end{tabular}
\end{table}

\begin{table}
\caption{
Atomic coordinates and thermal displacement parameters of IrTe$_2$ at 300 and 20 K. 
}
\begin{tabular}{lllll}
\hline
\hline
\multicolumn{5}{c}{300 K}\\
\hline
site & $x/a$ & $y/b$ & $z/c$ & $U_{\rm iso}$\\%
Ir1 & 0 & 0 & 0 & 0.01047(7) \\%
Te1 & 1/3 & 2/3 & 0.25305(4) & 0.01110(7) \\%
\hline
\hline
\multicolumn{5}{c}{20 K}\\%
\hline
site & $x/a$ & $y/b$ & $z/c$ & $U_{\rm iso}$\\%
Ir1 &   0 &  0 &  0 & 0.00356(6) \\%
Ir2 &    0.36069(5) &   0.21315(3) &   0.203394(15) & 0.00357(6) \\%
Ir3 &    $-$0.29000(6) & 0.43031(4) & 0.411623(16) &  0.00360(6)\\%
Te1 &    $-$0.00026(11) &  $-$0.05426(7) & 0.18477(3) & 0.00381(7) \\%
Te2 &   0.36127(11) & 0.27091(6) & 0.01679(3) & 0.00377(7) \\%
Te3 &   0.35701(10) & 0.15939(6) & 0.38782(3) &  0.00385(7) \\%
Te4 &  $-$0.27774(11) & 0.48038(7) & 0.22267(3) &  0.00386(7) \\%
Te5 &    0.08129(10) & 0.70076(6) & 0.41086(3)  &  0.00386(7) \\%
\hline
\hline
\end{tabular}
\end{table}

Figures 1(c) and 1(d) show the low-temperature crystal structure of IrTe$_2$ determined at 20 K. 
The structure is triclinic with the space group $P{\bar 1}$ ($\sharp$~2). 
The low-temperature structure is characterized by three crystallographically independent iridium sites, denoted as Ir1, Ir2, and Ir3 in Fig.~1(d). 
The bond lengths of adjacent iridium atoms along the trigonal $a$-axis are modulated considerably: 
The Ir3-Ir3 bond length of 3.069 {\AA} is considerably shorter than the Ir1-Ir2 and Ir2-Ir3 bond lengths (3.943 and 4.027 {\AA}, respectively), 
which results in a non-sinusoidal structural modulation along the trigonal $a$-axis that is in accordance with the electron diffraction measurements \cite{PhysRevLett.110.127209,PhysRevLett.108.116402}. 
The Ir3-Ir3 bond length is comparable to that of the dimerized Ir (3.012 {\AA}) of CuIr$_2$S$_4$ \cite{Nature416_155}. 
Thus, the structural phase transition of IrTe$_2$ can be viewed as the partial formation of a single Ir$_2$ dimer from the five iridium atoms along the trigonal $a$-axis. The average bond length is compatible with the average structure \cite{JLowTempPhys117_1129}. 
There is no structural modulation along the trigonal $b$-axis, and a uniform Ir bond length of 3.949 {\AA} is observed. 
Thus, along the trigonal $b$-axis direction there are stripes of dimerized Ir3  and non-dimerized Ir1 and Ir2. 
The position of these stripes shifts to $\pm{\bm a}$ along the trigonal $a$-axis in the adjacent layers, as shown in Fig.~1(c), 
resulting in a structural modulation of ${\bm q} = (1/5, 0, -1/5)$.

Modulation of the Te-Te distances appears in the triclinic phase, but it is much smaller than that of the Ir-Ir distances.
In the high-temperature trigonal phase, there are three characteristic Te-Te distances: 
3.498 {\AA} between adjacent IrTe$_2$ layers, 3.548 {\AA} between the upper and lower Te atoms within an IrTe$_2$ layer, and 
3.925 {\AA} between the in-plane Te atoms of an IrTe$_2$ layer. 
The short Te-Te bond between adjacent IrTe$_2$ layers has been discussed previously, and was shown to result in direct covalent Te-Te bonds and a Te $5p$ contribution to the electronic density of states at the Fermi level \cite{Jobic1992169}.
In the low-temperature phase,
a large modulation of 3.429--4.115 {\AA} was observed in the in-plane Te-Te distance, 
while modulations in the other Te-Te distances were smaller (3.368--3.547 {\AA} between adjacent IrTe$_2$ layers and 3.488--3.561 {\AA} between the upper and lower Te atoms in an IrTe$_2$ layer). 
The shortest Te-Te bond length of 3.368 {\AA} in the triclinic phase of IrTe$_2$ is much longer than that observed in the modulated structure of AuTe$_2$ with a distorted CdI$_2$-type structure (2.88 {\AA}) \cite{Schutte:bx0246,JPSJ.82.063704}.  
The smaller modulations of the Te bond lengths suggest a dominant Ir $5d$ contributes to the structural-electronic phase transition mechanism of IrTe$_2$, although there is a finite Te $5p$ contribution through the large mixing between the Ir $5d$ and Te $5p$ bands. 

\begin{figure}
\begin{center}
\includegraphics[width=6.5cm]{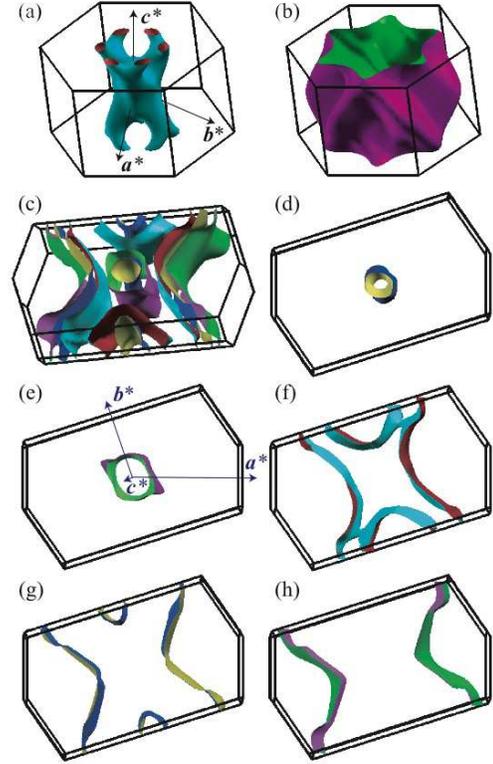}
\caption{(Color online) 
Fermi surfaces calculated for the [(a) and (b)] high-temperature trigonal phase and [(c)-(h)] low-temperature triclinic phase of IrTe$_2$.  
All the Fermi surfaces in the low-temperature phase are summarized in (c) and are illustrated separately in (d)-(h). 
The reciprocal lattice vectors $(\bm{a}^*,\bm{b}^*,\bm{c}^*)$ are shown in (e).  
In each Fermi-surface sheet, the contributions come predominantly from 
(d) Te3 and Te5; 
(e) Te1, Te2, and Te4; 
(f) Ir1, Ir2, Te1, Te2, and Te4; 
(g) Ir1, Ir2, Te1, Te2, and Te4; and 
(h) Ir3 and Te5.  
}
\end{center}
\end{figure}

\begin{figure}
\begin{center}
\includegraphics[width=7.5cm]{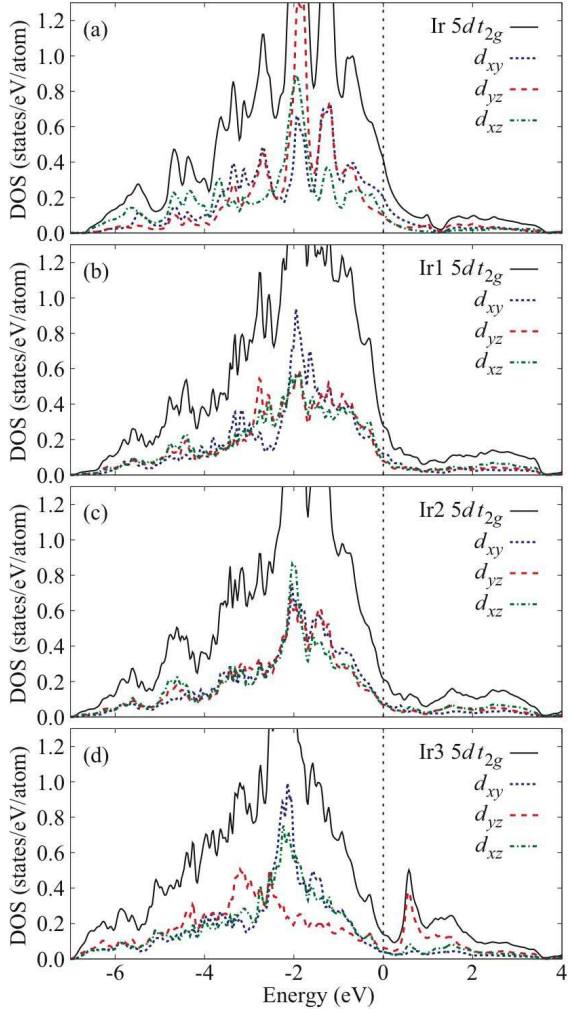}
\caption{(Color online) 
Orbital-decomposed partial DOS calculated for (a) the high-temperature trigonal phase and (b)-(d) the low-temperature triclinic phase of IrTe$_2$.  The $d_{xy}$, $d_{yz}$, and $d_{xz}$ components in the $t_{2g}$ manifold of the Ir $5d$ bands are shown for (b) Ir1, (c) Ir2, and (d) Ir3.  The vertical line represents the Fermi level.  
}
\end{center}
\end{figure}

To study the change in the electronic states across the structural phase transition, we carried out electronic structure calculations using the full-potential augmented plane-wave and local orbital methods, as implemented in the WIEN2k code \cite{WIEN2K}.  
We applied the generalized gradient approximation for electron correlations and considered the exchange-correlation part of the potential using the Perdew-Burke-Ernzerhof exchange-correlation functional \cite{PhysRevLett.77.3865}.  
Spin-orbit coupling (SOC) was taken into account for both Ir and Te.  
The maximum modulus of the reciprocal vectors $K_{\rm max}$ and muffin-tin radii of the atoms $R_{\rm MT}$ were chosen such that $R_{\rm MT}K_{\rm max}=7$.  
The lattice parameters and atomic coordinates were taken from our experimental data, and the Brillouin zone of the trigonal phase (triclinic phase) was sampled using a $16\times 16\times 10$ ($14\times 8\times 3$) $\bm{k}$ mesh.  
The calculated band dispersion, width, and Fermi surfaces of the high-temperature trigonal phase agree with those determined by the angle-resolved photoemission spectroscopy,\cite{PhysRevB.86.014519} indicating that the Hubbard-type repulsive interaction $U$ is negligible in IrTe$_2$. The finite $U$ in the GGA+$U$ \cite{Anisimov,Lichtenstein} scheme does not affect the basic features of the state near the Fermi level, thus the shape of the Fermi surface. 
No spin-polarized solutions were found, consistent with the absence of Curie-Wiess behavior.\cite{JPSJ.81.053701} 

Figure 3 shows the calculated Fermi surfaces of the high-temperature trigonal and low-temperature triclinic phases of IrTe$_2$.  
In the high-temperature phase, 
shown in Figs.~3(a) and (b), the Fermi surfaces are rather three dimensional and 
electronic quasi-two-dimensionality occurs in the basal $ab$ plane of the {\it trigonal} lattice [see Figs.~1(a) and (b)]. 
In the low-temperature phase, however, it is immediately clear in Figs.~3(c)-(h) that all five Fermi-surface sheets exhibit a fair quasi-two-dimensionality consisting of small cylinder-like surfaces and strongly warped quasi-one-dimensional-like sheets, the direction of which indicates that the quasi-two-dimensional conducting planes are now parallel to the $ab$ plane of the {\it triclinic} lattice [see Figs.~1(c) and (d)] or normal to the structural modulation vector $\bm{q}=(1/5,0,-1/5)$.  
This change in the Fermi surfaces demonstrates that switching of the conducting planes occurs from the basal plane of the trigonal IrTe$_2$ to the tilted plane normal to $\bm{q}$ of the triclinic IrTe$_2$
as well as the enhancement of two-dimensionality across the structural phase transition at approximately 270 K. 
The Fermi surfaces calculated using the average monoclinic structure with uniform Ir-Ir bonds for the low-temperature phase exhibit neither switching of conducting planes nor dimensional crossover.\cite{ScientificReports3.1153,PhysRevB.87.180501} 
Thus the partial formation of Ir dimers is crucial in switching the conducting planes and enhancing two dimensionality in the low-temperature triclinic phase of IrTe$_2$, reminiscent of the formation of As dimers in iron-based superconductors, which results in a dimensional crossover.\cite{PhysRevB.83.060505,JPSJ.80.103701}

Ir and Te are subject to strong SOC; the band exhibits a large spin-orbit (SO) splitting of the order of 1 eV.\cite{PhysRevB.87.180501} 
However, the SOC seems not to be involved in the mechanism of the structural phase transition of IrTe$_2$.  
Figure 4 shows the orbital-decomposed partial DOS calculated for the high-temperature trigonal and low-temperature triclinic phases of IrTe$_2$, where the $d_{xy}$, $d_{yz}$, and $d_{xz}$ components in the $t_{2g}$ manifold of the Ir $5d$ bands are illustrated [see Figs.~1(b) and (d) for the definition of the coordinate axes $(x,y,z)$]. 
Compared with the partial DOS for the high-temperature phase, a striking change occurs in the $d_{yz}$ component in the $t_{2g}$ manifold of the dimerized Ir3 atoms [Fig.~4(d)].  
The three $t_{2g}$ components are mostly occupied by electrons in the high-temperature phase [Fig.~4(a)], 
but the formation of Ir$_2$ dimers in the low-temperature phase results in strong bonding-antibonding splitting in the $d_{yz}$ bands of the two Ir3 atoms, which increases the energy of the antibonding bands above the Fermi level [note the peak at $\sim 0.7$ eV in Fig.~4(d)] and decreases the energy of the bonding bands by $\sim 1$ eV [note the peaks at $\sim -3$ eV in Fig.~4(d)]. 
The splitting is by far larger than the SO splitting,\cite{PhysRevB.87.180501} suggesting that lowering in energy of the $d_{yz}$ bonding orbital plays an important role in the mechanism of the structural phase transition of IrTe$_2$.\cite{SM} 
This is analogous to CuIr$_2$S$_4$, in which orbital ordering takes place at the structural phase transition.\cite{Nature416_155,mizokawa} 
The weight of the DOS is thus largely depleted around the Fermi level at the dimerized Ir3 sites.  
The partial DOS for the $p_z$ component of the $5p$ bands of Te5 located between the dimerized Ir3 atoms also undergoes a similar change near the Fermi level (results not shown here).
Thus total DOS is reduced from 1.89 to 0.97 states/eV/IrTe$_2$, consistent with the reduction of magnetic susceptibility and electronic specific heat.\cite{JLowTempPhys117_1129,ScientificReports3.1153,JPSJ.81.053701}  
The local DOSs at the Ir3 and Te5 sites are thus strongly reduced at the Fermi level, although a band gap does not appear, indicating that the dimerized Ir$_2$ planes become less conducting or effectively disconnect the conducting planes.  
In contrast, the planes in which the Ir1 and Ir2 atoms are located remain highly conducting as the height of the local DOS at the Fermi level suggests, yielding the quasi-two-dimensional conducting planes normal to the modulation vector $\bm{q} = (1/5, 0, -1/5)$.  
Such switching of the conducting planes that occurs with the structural phase transition is, to the best of our knowledge, quite a rare example in transition-metal compounds.  

In summary, we have employed single-crystal X-ray diffraction to study the structural-electronic phase transition of IrTe$_2$ at approximately 270 K.  
The low-temperature triclinic structure (space group $P\bar{1}$) is characterized by the partial formation of Ir$_2$ dimers that occurs in the triangular lattice of IrTe$_2$, resulting in a structural modulation with a wave vector of ${\bm q} = (1/5, 0, -1/5)$.  
First-principles band calculations have demonstrated that tilted two-dimensional Fermi surfaces emerge in the triclinic phase of IrTe$_2$, suggesting that  switching of the conducting planes occurs from the basal plane of the trigonal IrTe$_2$ to the tilted plane normal to ${\bm q}$ of the triclinic IrTe$_2$.  

Part of this work was performed at the Advanced Science Research Center, Okayama University. 
The study was partially supported by a Grant-in-Aid for Scientific Research (C) (No.~25400372) and for Young Scientists (B) (No.~24740238) from the Japan Society for the Promotion of Science (JSPS) and the Funding Program for World-Leading Innovation R{\&}D on Science and Technology (FIRST Program) from JSPS. The synchrotron radiation experiments performed at SPring-8 were supported by the Japan Synchrotron Radiation Research Institute (JASRI; Proposal Nos.~2011B1072, 2012B1463, 2012B1055).  
T.T. acknowledges support from the JSPS Research Fellowship for Young Scientists.

{\small 
\medskip\noindent
{\it Note added in proof} -- We noticed a paper by Pascut et al. [arXiv:1309.3548, Phys. Rev. Lett. (in press)], reporting similar results.  
}

\end{document}